# Layer-Hybridized Wigner Crystals in $MoSe_2/WS_2$ Moiré Superlattice

Tianyi Ouyang[1#], Yuze Meng[1#], Li Yan[1#], Yuxuan Chen[1#], Shuai Zhang[1], Xinyue Chen[1], Melike Erdi[2], Takashi Taniguchi[3], Kenji Watanabe[4], Seth Ariel Tongay[2], Benjamin Hunt[1], Ming Xie[5], Yong-Tao Cui[6,7], Su-Fei Shi[1*]

1. Department of Physics, Carnegie Mellon University; Pittsburgh, PA 15213, USA
2. School for Engineering of Matter, Transport and Energy, Arizona State University; Tempe, AZ 85287, USA
3. Research Center for Materials Nano architectonics, National Institute for Materials Science; 1-1 Namiki, Tsukuba 305-0044, Japan
4. Research Center for Electronic and Optical Materials, National Institute for Materials Science; 1-1 Namiki, Tsukuba 305-0044, Japan
5. Department of Physics, The University of Texas at Dallas, Richardson, TX 75080, USA
6. Department of Physics and Astronomy, University of California at Riverside; Riverside, CA 92521, USA
7. Department of Materials Science and Engineering, University of California at Riverside; Riverside, CA 92521, USA

[#] These authors contributed equally to this work
[*] Corresponding author: sufeis@andrew.cmu.edu

## Abstract

**Transition metal dichalcogenide moiré heterobilayers with type-II band alignment provide a versatile platform for layer-polarized generalized Wigner crystals, in which strong Coulomb interactions drive charge ordering at fractional lattice fillings. With a finite interlayer band offset, an out-of-plane electric field can tune layer-resolved moiré bands through resonance and enable controllable interlayer hybridization. Although hybridized Mott insulators have been previously demonstrated, whether fractional charge-ordered states can survive such hybridization remains elusive. Here we drive an H-stacked $MoSe_2/WS_2$ moiré heterobilayer through a type-I-to-type-II band-alignment transition and realize layer-hybridized Mott insulator and generalized Wigner crystals. For fillings below one electron per moiré cell, tunneling delocalizes electrons and modifies Wigner crystallization. However, above one electron per cell, Coulomb repulsion overcomes tunneling and favors layer-separated occupation, stabilizing stronger charge-ordered states. These results establish electrically tunable hybridized moiré heterobilayers as a powerful platform for engineering correlated charge order and exploring fractional Chern phases and emergent magnetism.**

Transition metal dichalcogenide (TMDC) moiré superlattices provide a versatile platform for exploring strongly correlated quantum phases[1-11]. Among these systems, type-II heterobilayer moiré superlattices have been studied extensively[3,5,7,10-14], with electrons and holes residing in different layers. When the interlayer band offset is finite, an out-of-plane electric field provides an additional tuning parameter, enabling two layer-resolved flat bands to be brought into resonance and allowing controllable interlayer tunneling. Layer-hybridized Mott insulators have previously been realized[15-19] and the one in the $MoTe_2/WSe_2$ moiré bilayer has been shown to exhibit the anomalous quantum Hall effect. Generalized Wigner crystal (GWC) states[3,7,13,20], however, rely on longer-range Coulomb interactions and are generally more fragile than Mott insulators. It therefore remains unclear whether such charge-ordered states can survive interlayer hybridization. At the same time, layer-hybridized generalized Wigner crystals could host electrically tunable lattice symmetries, including emergent honeycomb lattices[19,21,22], and provide a pathway toward topological correlated states for fractional lattice fillings[6,18,23-25].

In this work, we use $MoSe_2/WS_2$ as an archetypal platform to investigate these questions. The conduction-band minima (CBMs) of $MoSe_2$ and $WS_2$ are close in energy, and an electrically driven transition between type-I and type-II band alignment has previously been demonstrated[26,27]. Moiré coupling arising from the lattice mismatch produces flat bands that host correlated electronic states. In particular, at a filling of one electron per moiré unit cell, the $MoSe_2/WS_2$ system exhibits magnetism in the type-I configuration[28,29], where electrons reside in the $MoSe_2$ layer. Interlayer hybridization may therefore provide a route toward emergent frustrated magnetic states[30].

Here, we investigate the H-stacked $MoSe_2/WS_2$ heterobilayer, in which the CBMs of $MoSe_2$ and $WS_2$ share the same spin configuration. We apply an out-of-plane electric field to tune the two bands into resonance and enable interlayer hybridization. We monitor the redistribution of doped electrons between the $MoSe_2$ and $WS_2$ layers using doping-dependent reflectance contrast spectroscopy and observe moiré-modified intralayer excitons in the $MoSe_2$ and $WS_2$ layers[31-35]. Using the 2s exciton of a nearby $WSe_2$ monolayer as a sensor[13,36], we find that GWC states persist as the CBMs of $MoSe_2$ and $WS_2$ are tuned into the interlayer-hybridization regime. Their evolution in this regime, however, depends strongly on filling. For the filling factor n between 0 and 1, interlayer tunneling delocalizes electrons across the two layers and modifies the GWC, as the increased kinetic energy reduces correlation strength. For the filling factor n larger than 1, electrons on doubly occupied sites preferentially occupy the opposite layer to minimize Coulomb repulsion. This interaction-driven layer separation stabilizes the GWC states and enables the observation of correlated insulator states that would otherwise be conductive. Electric-field-dependent magnetic circular dichroism (MCD) measurements further reveal that the layer-hybridized Mott insulator at one electron per moiré cell ($n = 1$) inherits the magnetic properties of the $n = 1$ state of $MoSe_2/WS_2$ in the type-I alignment. These results show how the competition between interlayer tunneling and Coulomb repulsion reorganizes correlated states across different fillings, establishing an electrically tunable platform for engineering layer-selective charge and magnetic order. More broadly, they inspire a route toward realizing and controlling complex many-body phases in hybridized moiré systems.

## Mott insulator and generalized Wigner crystal states in H-stacked $MoSe_2/WS_2$

A typical dual-gated device is schematically illustrated in Fig. 1a, in which the carrier density and out-of-plane electric field are independently controlled by combinations of the top- and back-gate voltages. A monolayer $WSe_2$, separated from the $MoSe_2/WS_2$ heterostructure by an h-BN spacer of about 2 nm in thickness, serves as a remote conductivity-sensing layer[13]. The twist angle of $MoSe_2$ and $WS_2$ is approximately 58.7°, corresponding to H-type stacking (60° stacking). Due to the moiré coupling, the conduction bands of $MoSe_2$ and $WS_2$ are both flat[20,36], with kinetic energy suppressed.

As we electrostatically introduce electrons to the moiré bilayer, various correlated insulating states, including the Mott insulator state (n = 1) and GWC at fractional fillings, are formed, as evidenced by the reflectance contrast spectra from the 2s Rydberg excitons of the $WSe_2$ sensing layer[13] (Fig. 1e). The $MoSe_2/WS_2$ heterobilayer is expected to have a type I alignment, and the CBM of $MoSe_2$ is lower than that of $WS_2$. As a result, the doping-introduced electrons first populate the $MoSe_2$ layer. This is confirmed by the reflectance contrast spectra of intralayer excitons in the $WS_2$ layer (Fig. 1d), as they remain consistent with those of charge-neutral excitons, up to the electron-doping shown in Fig. 1d. This is also confirmed by the reflectance contrast spectra of intralayer excitons in the $MoSe_2$ layer, as shown in Fig. 1c. For the charge-neutral region (around filling factor n = 0), the reflectance contrast spectra show moiré intralayer excitons $\mathrm{M}_1$ and $\mathrm{M}_2$, arising from moiré potential modulation of the original intralayer exciton in $MoSe_2$ (the higher-energy moiré exciton $\mathrm{M}_3$ is not shown here due to overlap with $WSe_2$ excitons, see Supplementary Information Section 5 for details)[31]. In the electron-doping region between filling factor n = 0 and n = 2, a trion peak emerges as $\mathrm{M}_1^-$ due to the electron doping in $MoSe_2$[29,32].

The intensity of the moiré intralayer trion $\mathrm{M}_1^-$ is reported to be a sensitive probe of the electron occupation in the $MoSe_2$ layer[32]. As shown in Fig. 1c, the intensity of $\mathrm{M}_1^-$ increases as $\mathrm{n}_{\mathrm{Mo}}$ (the electron filling factor in the $MoSe_2$ layer) increases from 0 to 1 and decreases as $\mathrm{n}_{\mathrm{Mo}}$ further increases from 1 to 2. It eventually vanishes when the $MoSe_2$ moiré site becomes doubly occupied (Extended Fig. 1).

If the $MoSe_2$ or $WS_2$ layer is electron-doped to half-filling, i.e., one electron per moiré cell, the CBM of either layer will be split into lower and upper Hubbard bands (LHB and UHB, Fig. 1b). The above observations suggest that, for 1 < n < 2, the electrons populate the UHB of $MoSe_2$ ($\mathrm{UHB}_{\mathrm{Mo}}$), which is lower in energy than the CBM of $WS_2$ in the absence of the out-of-plane electric field.

## Electrically tunable and layer-hybridized conduction bands

We next explore the doping-dependent reflectance contrast spectra under different electric fields to investigate the relative conduction band alignment between $MoSe_2$ and $WS_2$. Fig. 2a shows the $MoSe_2$ reflectance contrast spectra under an electric field of 0.14 V/nm. In contrast to the zero-electric field case, the $\mathrm{M}_1^-$ peak intensity increases as the total filling rises from n = 0 to 1 but remains at its maximum peak intensity from n = 1 to 2 (extracted intensity shown in Fig. 2d, purple dots). Meanwhile, under the same electric field, the $WS_2$ exciton remains neutral-exciton response below n = 1 but exhibits attractive and repulsive exciton-polaron responses ($\mathrm{W}^-$ and $\mathrm{W}^+$) for n > 1, as shown in Fig. 2b. These observations indicate that the first electron occupies the $MoSe_2$ layer, whereas the second electron occupies the $WS_2$ layer, with the LHB of $WS_2$ ($LHB_W$) lower in energy than the $UHB_{Mo}$ (Schematically shown in Fig. 2c).

By further increasing the electric field to 0.18 V/nm, as shown in Fig. 2i, the $\mathrm{M}_1^-$ response disappears below n = 1. However, it re-emerges and reaches a maximum at n = 2 (Fig. 2l, blue dots). Meanwhile, the $WS_2$ layer exhibits $\mathrm{W}^-$ and $\mathrm{W}^+$ responses at all finite electron fillings, as shown in Fig. 2j. These observations indicate that the first electron occupies the $WS_2$ layer, whereas the second electron occupies the $MoSe_2$ layer, with the LHB of $MoSe_2$ ($LHB_{Mo}$) lying between the $LHB_W$ and $UHB_W$ (Upper Hubbard band of $WS_2$), schematically shown in Fig. 2k. In this configuration, the $MoSe_2/WS_2$ heterostructure has a type-II band alignment as the valence band maximum (VBM) remains in $MoSe_2$ layer due to a large band offset.

Interestingly, at an intermediate electric field of 0.16 V/nm, the $\mathrm{M}_1^-$ peak is present at n = 0 and the intensity remains strong up to n = 2, as shown in Fig. 2e (extracted intensity shown in Fig.2h, red dot). Concurrently, the $WS_2$ layer spectra show both charge-neutral exciton and attractive polaron $\mathrm{W}^-$ (or trion) responses between n = 0 and 1, which evolve into $\mathrm{W}^-$ and $\mathrm{W}^+$ above n = 1. More intriguingly, despite the presence of the $\mathrm{W}^-$ in $WS_2$ between n = 0 and 1, the repulsive exciton polaron $\mathrm{W}^+$ in $WS_2$ does not appear until n > 1, different from the case of the higher electric field of 0.18 V/nm in which the electron doping immediately leads to the simultaneous emergence of both $\mathrm{W}^-$ and $\mathrm{W}^+$. We interpret these observations as the formation of a hybridized band between the CBMs of $WS_2$ and $MoSe_2$, which share the same spin configuration in the H-stacked structure. Consequently, for 0 < n < 1, the electrons are shared between the $MoSe_2$ and $WS_2$ layers.

To investigate this hybridization, we extract the $\mathrm{M}_1^-$ peak intensities as a function of doping level over the electric-field range of 0.14-0.18 V/nm (See Extended Data Fig. 1). At n = 1, the $\mathrm{M}_1^-$ peak intensity initially remains at its maximum and then drops in the intermediate-field regime. We fit the field dependence using a two-level hybridization model, assuming that the $\mathrm{M}_1^-$ peak intensity is proportional to the probability of the hybridized electrons residing in $MoSe_2$. The model reproduces the observed electric-field evolution and yields a hybridization energy (i.e., interlayer coupling strength) of $t = 1.95 \pm 0.42\ \mathrm{meV}$. This hybridization energy is of the same order as the flat band width but is smaller than the previously reported hybridization of the $WSe_2$ valence band[15,16] (10-30 meV) and hybridization of the $WS_2$ conduction band (~ 6 meV). The small value of $t$ might arise from the fact that the hybridized electrons in $MoSe_2$ and $WS_2$ are localized in different high-symmetry points[32,33], reducing the tunneling rate.

The hybridization energy is much smaller than the on-site repulsion energy in this moiré system. Following our previous discussion of Fig. 2, we also investigate the reflectance contrast spectra as a function of electric field for doping n = 1 and 2 (see Supplementary Information Sections 3 and 4 for details). By analyzing the electric-field-driven alignment of $LHB_{Mo}$ and $UHB_{Mo}$ relative to the $LHB_W$ and $UHB_W$ for the correlated states at n=1 and 2, we determine the Hubbard band splitting in $MoSe_2$ and $WS_2$ layers to be $U_{Mo}$ = 55 meV and $U_W$ = 22 meV, respectively, consistent with the values reported in other works[29,35]. We also find the energy offset at zero electric field between the CBMs of $MoSe_2$ and $WS_2$ to be $\Delta_{offset}$ = 113 meV in this H-stacked configuration, similar to the value reported for non-aligned $MoSe_2/WS_2$ configuration[36].

## Interlayer-hybridized generalized Wigner crystal states

We further investigate the correlated states such as Mott insulator and, more interestingly, GWC states in the hybridization regime. Applying different out-of-plane electric fields to realize the band alignments corresponding to schematics in Fig. 2c, g, k, we measure $WSe_2$ sensing spectra as a function of doping in the $MoSe_2/WS_2$ moiré bilayer, as shown in Fig. 3a-c. It is evident that most of the correlated states survive in the hybridization regime (Fig. 3b), compared with the scenario in which electrons reside only in the $MoSe_2$ (Fig. 3a) or $WS_2$ layer (Fig. 3c). However, the fractional filling states in the range $0 < n < 1$ are less pronounced while the ones in the range $1 < n < 2$ are more pronounced.

For the lower-energy branch (bonding branch) of the hybridized CBM, the Hubbard model would give rise to LHB and UHB, and full filling of the LHB (i.e., half filling of the bonding branch) leads to the Mott insulator at n = 1. We note that this is a hybridized Mott insulator, similar to those reported in other moiré systems[15,17,18], as the electron is shared between the $MoSe_2$ and $WS_2$ layers, but maintains a density of one electron per moiré cell. For $n < 1$, the electron retains the hybridization character and is shared between the $MoSe_2$ and $WS_2$ layers, illustrated in Fig. 3f (upper panel). This hybridization increases kinetic energy and leads to reduced correlation strength, leading to less pronounced features in the corresponding 2s sensing spectra (Fig.3b). It is worth noting that the n = 2/3 state is no longer an insulating state, while the n = 1/2 remains a GWC state, which is surprising since n = 1/2 is expected to have a smaller bandgap than the n = 2/3 state in a non-hybrid regime[13,20,37,38]. We attribute this to increased Coulomb screening due to the extended wavefunction of the hybridized state, while the exact microscopic mechanism needs to be explored in the future.

However, n = 2 is not a hybridized Mott insulator. Since the hybridization is much less than the repulsion energy in the $MoSe_2$ and $WS_2$ layers, the two electrons will be separated into two layers, with one electron in $MoSe_2$ and one in $WS_2$ (Illustrated in Fig. 3f lower panel). This layer separation is similar to the two-exciton layer separation that forms a staggered dimer in a trilayer moiré system[16]. This layer separation also leads to a more pronounced GWC state in the hybridization regime for the filling factor n between 1 and 2. For n larger than 1, any cell with two electrons will have them layer-separated and be more confined in each moiré site of $MoSe_2$ and $WS_2$ (Illustrated in Fig. 3f middle panel). The two-electron correlated states (see Supplementary Information Section 8), with reduced kinetic energy, are different from and have stronger correlation strength than the two single-electron pictures for $1<n<2$ corresponding to Fig. 3a and Fig. 3c, in which the first electron is in the $MoSe_2$ layer and the second in the $WS_2$ layer (Fig. 3a) or vice versa (Fig. 3c). As a result, the GWC states are strengthened due to increased correlation strength, and we even observe GWC of n = 3/2 in the hybridization regime, which is missing in the scenario without hybridization. The absence of the correlated insulator states at 2/3 for $0 < n < 1$ and the appearance of the insulating state of 3/2 for $1 < n < 2$ can be clearly seen in the color plot Fig. 3d and its line traces in Fig. 3e.

**Electric-field-tunable magnetism**

Finally, the hybridization between moiré flat bands of $MoSe_2$ and $WS_2$ provides a platform to investigate intriguing magnetism in the hybridized state. It has been shown that the n = 1 state of the $MoSe_2/WS_2$ moiré bilayer, when electrons occupy the $MoSe_2$ layer with the density of one electron per moiré cell, is ferromagnetic in the R-stacked configuration and antiferromagnetic in the H-stacked configuration[28,29]. Meanwhile, the n = 1 state in the $WS_2/WSe_2$ moiré bilayer, when electrons reside in the $WS_2$ layer, is nonmagnetic. We would expect that the n = 1 state in the hybridization regime inherits magnetism to a certain degree. Fig. 4a shows the electric-field-dependent $MoSe_2$ reflectance contrast spectra at n = 1. The $\mathrm{M}_1^-$ peak disappears near 0.16 V/nm, corresponding to the hybridization regime. The detailed analysis of the electric-field-dependent spectra is provided in Supplementary Sections 2 and 3. Figures 4b–d show the MCD responses measured at electric fields of 0.20, 0.16, and 0 V/nm, as marked in Fig. 4a (See Methods for the definition of MCD). These three conditions correspond to the n=1 occupying the $MoSe_2$ layer only, shared between the $MoSe_2$ and $WS_2$ layers, and the $WS_2$ layer only, respectively. For the n = 1 state, when the electrons only reside in the $MoSe_2$ layer, the MCD signal exhibits clear saturation at high magnetic fields, as shown in Fig. 4d, consistent with the expectation of emerging magnetism. When the electrons reside only in the $WS_2$ layer, this saturation behavior is absent, as shown in Fig. 4b, consistent with a non-magnetic state[39]. In the hybridization regime, when the electric field is 0.16 V/nm, the saturation behavior reappears.

We quantify the magnetic response via the slope of the MCD spectra in the linear regime as $\chi_{MCD} = d(\mathrm{MCD})/dB|_{B=0}$, which is proportional to the out-of-plane magnetic susceptibility. In the initial type-I band alignment regime, the MCD susceptibility at 0 V/nm is estimated to be $\chi_{MCD}$ = 0.08% $T^{-1}$ (Fig. 4d). In the hybridization regime, for the n=1 state at 0.16 V/nm, MCD susceptibility is estimated to be $\chi_{MCD}$ = 0.03% $T^{-1}$ (Fig. 4c), while in the type-II regime at 0.20 V/nm, $\chi_{MCD}$ = 0.01% $T^{-1}$ (Fig. 4b). The reduced but finite magnetism of the hybridized Mott insulator is consistent with the hybridization picture.

## Conclusion

We use an out-of-plane electric field to drive a band alignment transition from type-I to type-II in a $MoSe_2/WS_2$ moiré heterostructure. In the intermediate electric-field range, we have identified a regime in which the two flat moiré bands hybridize, and correlated electrons are shared between two layers. These correlations lead to hybridized Mott and GWC states in which the ordered electron lattices are shared between two layers. These fascinating hybridized correlated electron states add to the knowledge and capability of exploring quantum many-body states. Considering the underlying honeycomb lattice, we are inspired to explore emerging topological states and frustrated magnetism from this system in the future.

## Methods

### Sample fabrication

TMDC moiré heterostructures were fabricated using the dry pick-up method described in our previous work[16]. The assembled heterostructure was released onto a substrate with pre-patterned electrodes at 170 °C. Polycarbonate (PC) residues were subsequently removed by sequential rinsing with chloroform and isopropanol. The device was then dried under a nitrogen flow and annealed at 250 °C for 8 h under vacuum (<$10^{-6}$ Torr).

### Optical characterization

All measurements were taken in an Attodry 2100 system at a temperature of 1.67 K. A home-built confocal microscope system was used to focus the white light on the sample and collect the optical signal into a spectrometer (Princeton Instruments). The reflectance contrast measurements were performed with a broadband white light source (Thorlabs, OSL2). A relative reflectance background $R_0$ was obtained from the reflectance spectrum at a high electron-doping level. The reflectance contrast is defined as $\Delta R/R = (R - R_0)/R_0$.

### Magnetic circular dichroism measurements.

Magneto-optical measurements were performed under an out-of-plane magnetic field. Circularly polarized light was generated using a linear polarizer and a quarter-wave plate. The reflectance spectra under left- and right-circularly polarized excitation, $R_+(E, B)$ and $R_-(E, B)$, respectively, were recorded under otherwise identical conditions. The magnetic circular dichroism was calculated as

$$\mathrm{MCD}(E, B) = \frac{R_+(E, B) - R_-(E, B)}{R_+(E, B) + R_-(E, B)}.$$

The reported MCD signal was obtained by averaging the MCD spectrum centered at the $\mathrm{M}_2$ exciton resonance, near 1.63 eV.

### Determination of the nominal electric field

We obtained the thickness of the top and bottom h-BN flakes by atomic force microscopy (AFM) and defined them as $d_1$ and $d_2$. The nominal electric field is defined as the electric field in heterostructures, which is given by $E = \frac{\epsilon_{BN}}{2\epsilon_{TMDC}}\left(\frac{V_{TG}}{d_1} - \frac{V_{BG}}{d_2}\right)$, where $V_{TG}$ ($V_{BG}$) is the top (bottom) gate voltage, $d_1$ ($d_2$) is the thickness of the top (bottom) BN flake as $d_1$=$d_2$=55 nm, $\epsilon_{BN} = 3.5$ and $\epsilon_{TMDC} = 7$ are the relative dielectric constants of h-BN and TMDC, respectively.

### Determination of twist angle

We determine the twist angle from the moiré density extracted from the doping-dependent reflectance contrast spectra. The voltage difference between n = 1 and n = 2 obtained from the $WSe_2$ 2s sensing spectra is ΔV = 7.7 V. Because this interval corresponds to adding one electron per moiré unit cell, the moiré density is calculated as $n_{moiré} = \frac{\varepsilon_0 \varepsilon_{BN} \Delta V}{e d_{BN}}$. $\varepsilon_0$ is the vacuum permittivity, and e is the elementary charge. This gives $n_{moiré}$ = 2.71 × $10^{12}$ $cm^{-2}$. The deviation θ from perfect H-type alignment is then obtained directly from $n_{moiré} = \frac{2(\delta^2 + \theta^2)}{\sqrt{3}a^2}$, where a ≈ 0.322 nm is the average lattice constant of $MoSe_2$ and $WS_2$ and δ ≈ 4.4% is the lattice mismatch. This gives θ ≈ 1.3°, corresponding to a twist angle of approximately 58.7°.

## Data availability

The source data for Figs. 1-4 are provided with this paper. All other data that support the plots within this paper and other findings of this study are available from the corresponding authors upon reasonable request.

## Acknowledgments

S.-F.S. acknowledges support from NSF (Career Grant DMR-1945420, DMR-2104902, ECCS-2344658, and ECCS-2139692). S.-F.S. also acknowledges the Gordon and Betty Moore Foundation, grant https://doi.org/10.37807/GBMF13836. Y.-T.C. acknowledges support from NSF under awards DMR-2104805 and DMR-2145735. K.W. and T.T. acknowledge support from the CREST (JPMJCR24A5), JST and World Premier International Research Center Initiative (WPI), MEXT, Japan. S.T. acknowledges support from DOE-SC0020653 (initial crystal to 2D excitonic benchmarking), Applied Materials Inc. (bulk crystal growth), and DMR 2330110 (defect/stability/ performance correlation).

## Author contributions

T.O., Y.M., and S.-F.S. conceived the project. Y.M. and X.C. fabricated heterostructure devices. T.O., L.Y., and Y.C. performed the optical spectroscopy measurements. M.E. and S.T. grew the TMDC crystals. T.T. and K.W. grew the h-BN crystals. S.-F.S., T.O., and Y.M. analyzed the data. S.-F.S. wrote the manuscript with the help of T.O. and Y.M., along with input from all authors.

These authors contributed equally: T.O., Y.M., L.Y., and Y.C.

## Competing interests

The authors declare no competing interests.

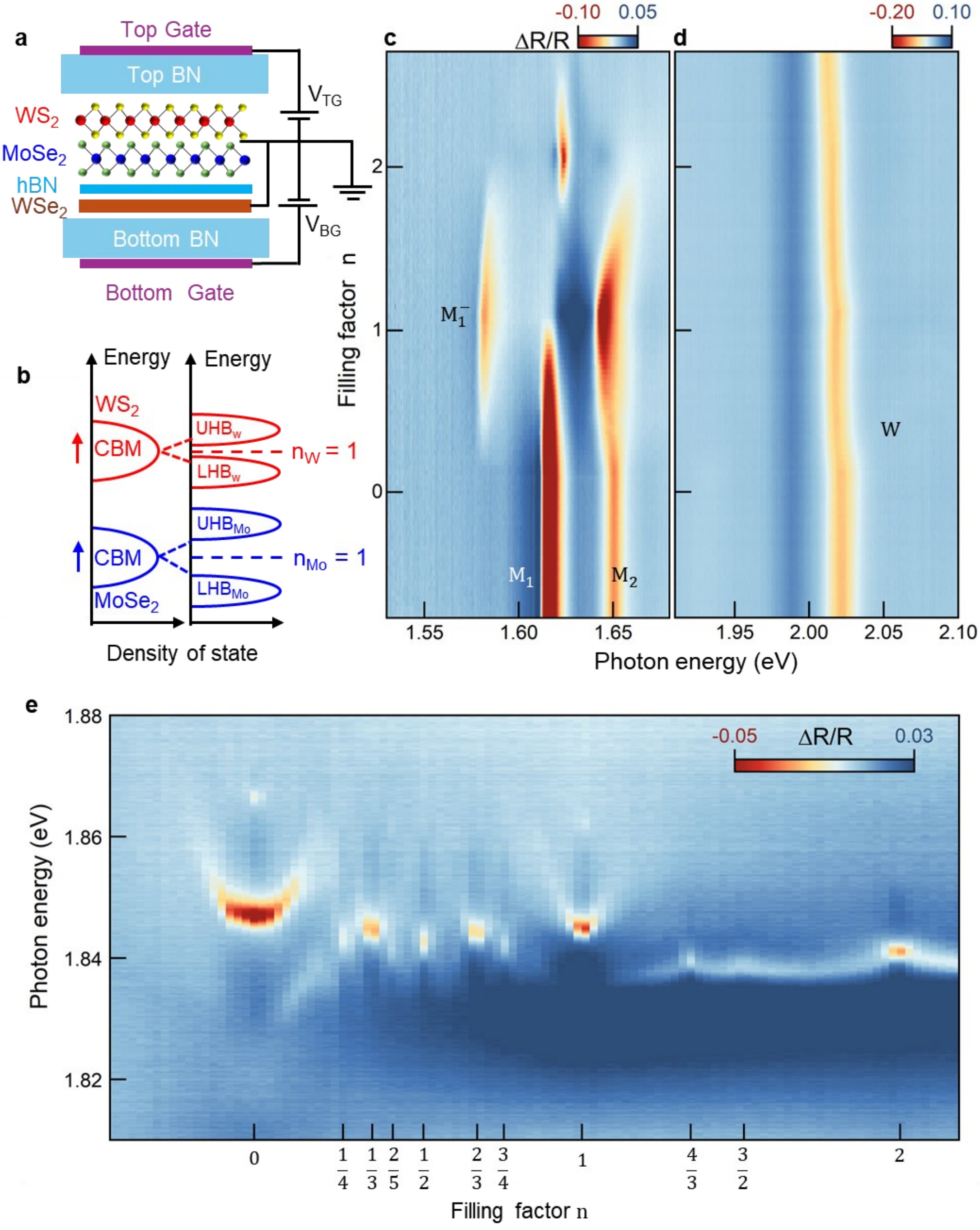

**Fig.1. Doping-dependent reflectance contrast spectra of the $MoSe_2/WS_2$ moiré heterobilayer at zero electric field.** (a) Schematic of the $MoSe_2/WS_2$ heterobilayer device incorporating a $WSe_2$ 2s exciton sensing layer. (b) Schematic of the Hubbard-band splitting of the conduction-band minima (CBMs) in $MoSe_2$ and $WS_2$. (c)–(e) Doping-dependent reflectance contrast spectra of $MoSe_2$, $WS_2$ and $WSe_2$ 2s sensing in the heterojunction region, respectively. M and W represent intralayer moiré excitons in $MoSe_2$ and $WS_2$, respectively. The sample temperature is maintained at 1.67 K for all measurements shown in this work.

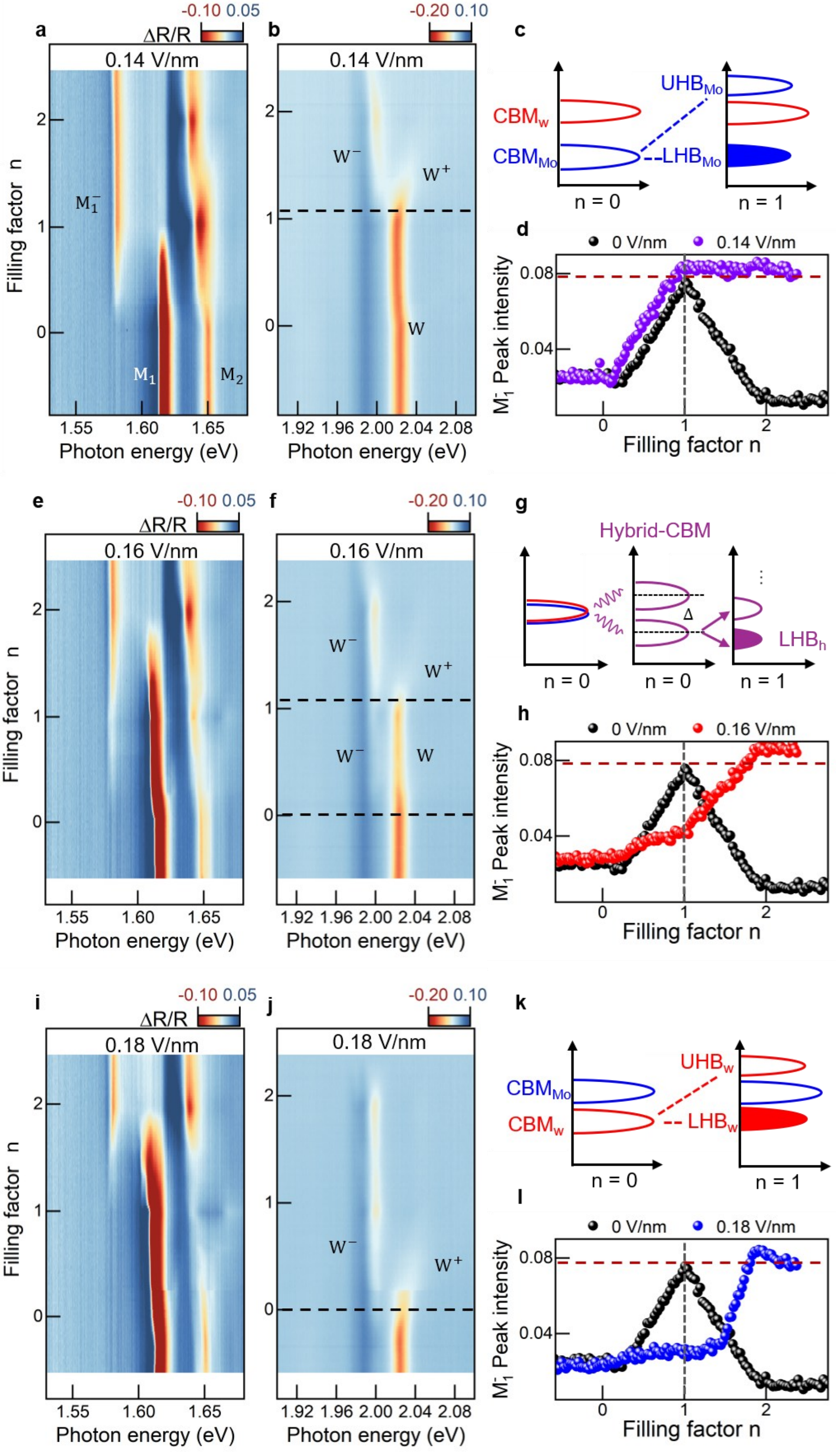

a
b
c
d
e
f
g
h
i
j
k
l
ΔR/R
-0.10 0.05
-0.20 0.10
0.14 V/nm
0.16 V/nm
0.18 V/nm
Filling factor n
Photon energy (eV)
M₁⁻
M₁
M₂
W⁻
W⁺
W
CBM_W
CBM_Mo
UHB_Mo
LHB_Mo
n = 0
n = 1
0 V/nm
M₁⁻ Peak intensity
Hybrid-CBM
Δ
LHB_h
UHB_W
LHB_W

**Fig.2. Electric-field-tuned band-alignment switching and flat-band hybridization.** (a), (b) Doping-dependent reflectance contrast spectra of $MoSe_2$ and $WS_2$, respectively, measured in the heterojunction region at an electric field of 0.14 V/nm. (c) Corresponding schematic illustration of the Hubbard-band alignment. (d) Corresponding extracted doping-dependent $M_1^-$ peak intensity. Red dashed line denotes the maximum intensity of $M_1^-$, while the grey dashed line denotes the filling factor n = 1. (e), (f) Doping-dependent reflectance contrast spectra of $MoSe_2$ and $WS_2$, respectively, at 0.16 V/nm. (g) Corresponding hybridized flat bands and Hubbard band splitting. Red/blue: flat CBM in $WS_2$/$MoSe_2$. Purple in middle: interlayer hybridized flat bands. Purple on the right: interlayer hybridized LHB and UHB. (h) Corresponding $M_1^-$ peak Doping-dependent intensity. (i), (j) Doping-dependent reflectance contrast spectra of $MoSe_2$ and $WS_2$, respectively, at 0.18 V/nm. (k) Corresponding Hubbard-band alignment. (l) Corresponding $M_1^-$ Peak doping-dependent intensity.

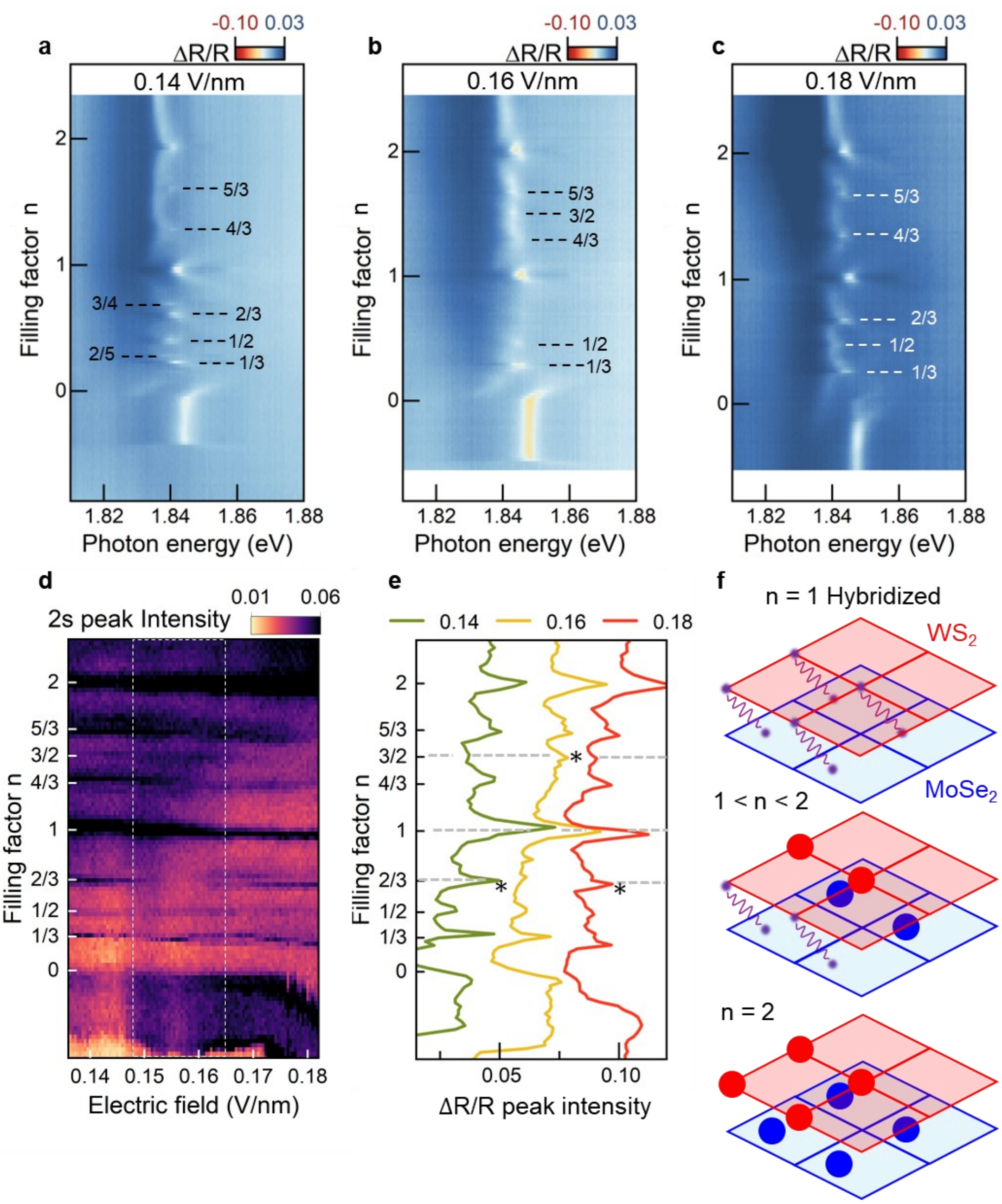

**Fig.3. $WSe_2$ 2s sensing of flat-band hybridization in the band-alignment transition regime.** (a)–(c) Doping-dependent reflectance contrast spectra of the $WSe_2$ 2s sensing exciton measured at electric fields of 0.14, 0.16, and 0.18 V/nm, respectively. (d) Large-range map of the $WSe_2$ 2s peak intensity across the band-alignment transition regime. White dashed box denotes the transition regime. (e) Doping-dependent $WSe_2$ 2s sensing exciton peak intensity at electric fields of 0.14, 0.16, and 0.18 V/nm. Stars denote the correlated insulating states at n = 2/3 and 3/2. (f) Schematic illustrations of hybridized Mott insulator n = 1 (upper panel), nonhybridized charge-transfer insulator n = 2 (lower panel), and a mixture of these two at 1 < n < 2. Red and blue spheres represent the charges in $WS_2$ (AB site) and $MoSe_2$ ($B_{S/Se}$ site) layers, respectively, while purple spheres represent hybridized charges distributed over two layers and two different moiré sites.

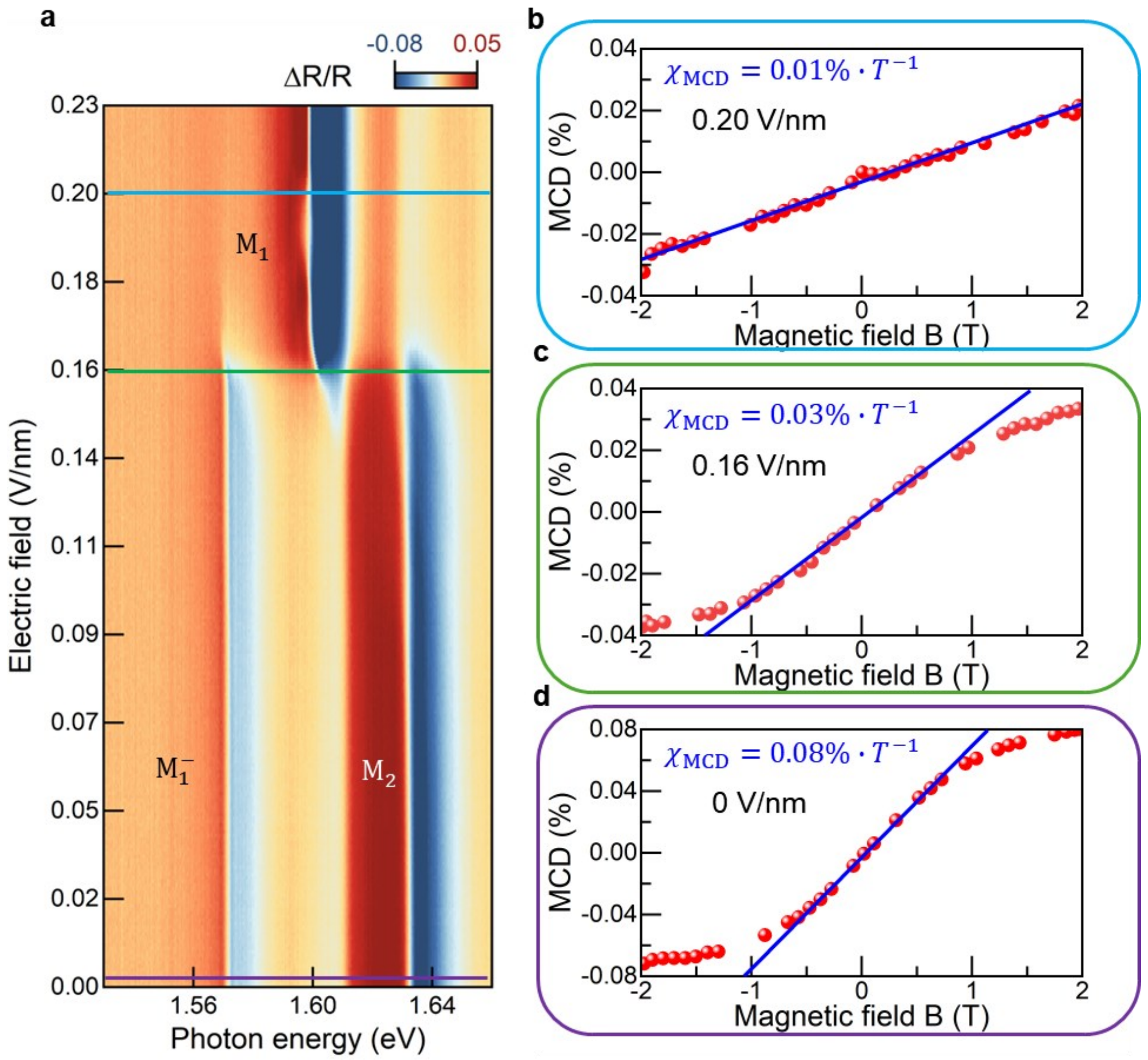

**Fig.4. Magnetic circular dichroism across the band-alignment transition.** (a) Electric-field-dependent reflectance contrast spectra of $MoSe_2$ in the junction region at n = 1. The blue, green, and purple solid lines mark representative electric fields in the type-II band-alignment regime, the transition regime, and the type-I band-alignment regime, respectively. (b)-(d) Magnetic-field-dependent MCD measurements at resonant energy of 1.63 eV, corresponding to the blue, green, and purple solid lines in (a), respectively. All blue lines in (b)-(d) are fitted near zero magnetic field.

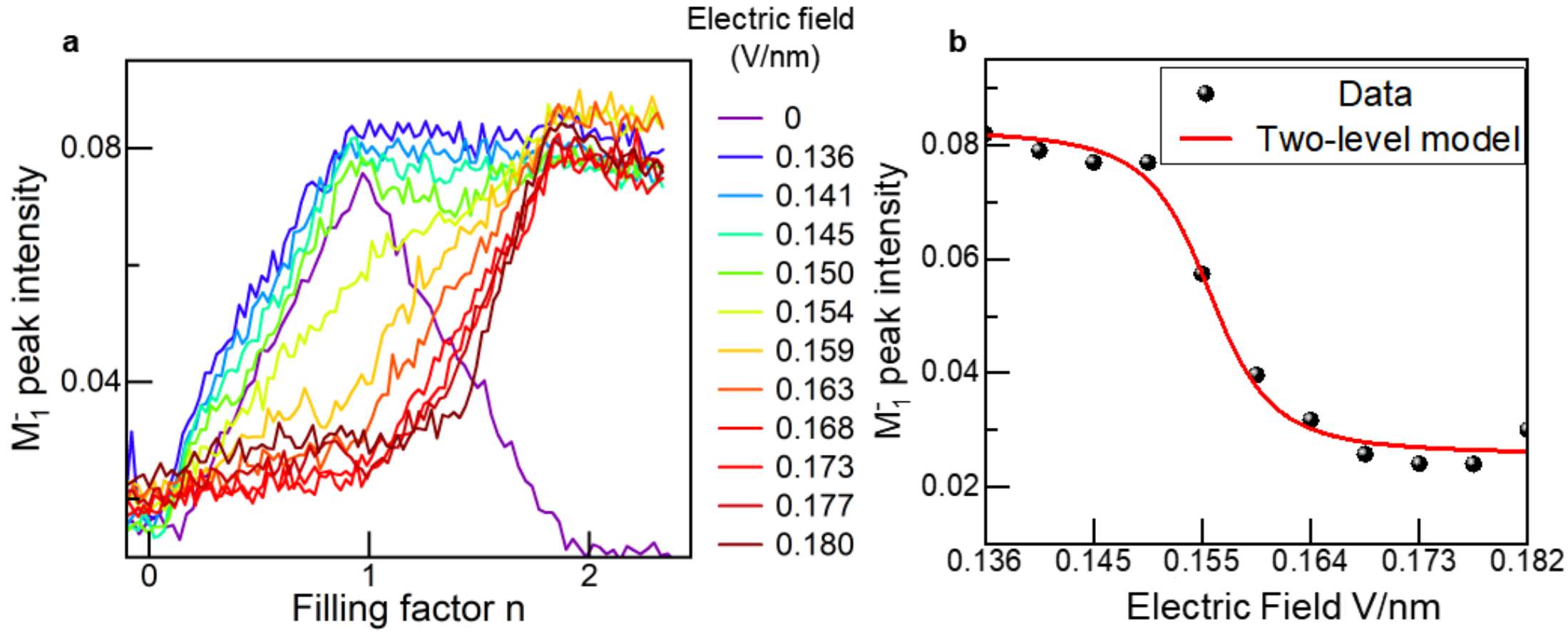

**Extended Data Fig.1 | Extracted $\mathrm{M}_1^-$ reflectance contrast peak intensity under various electric fields and two-level model fitting.** (a) Extracted $\mathrm{M}_1^-$ reflectance contrast peak intensity under various electric fields. (b) The $\mathrm{M}_1^-$ reflectance contrast peak intensities at n = 1 as a function of electric fields, fitted by a two-level model. Extracted hybridization energy $t = 1.95 \pm 0.42\ \mathrm{meV}$.

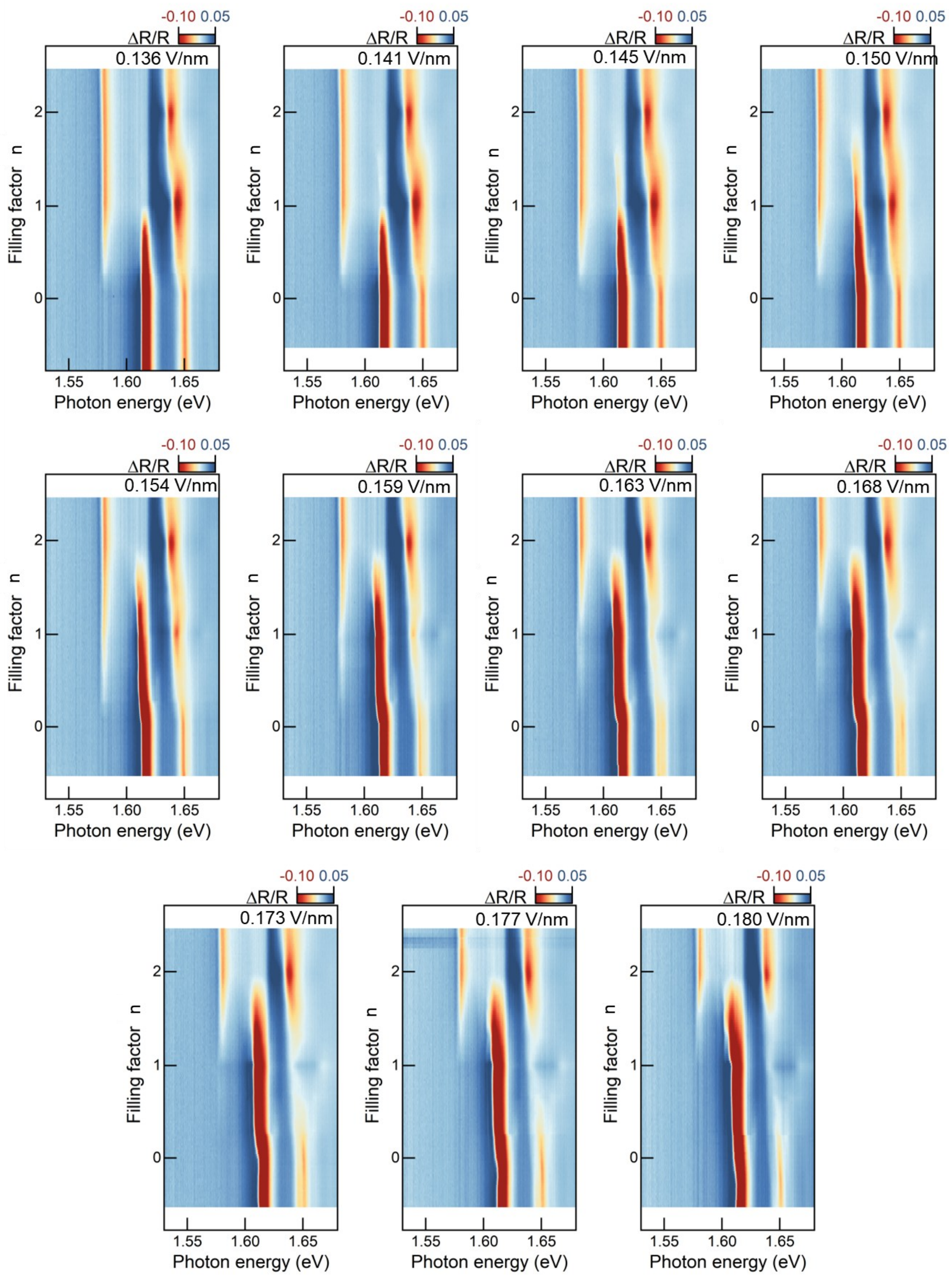

**Extended Data Fig.2 | Doping-dependent reflectance contrast spectra of $MoSe_2$ in the heterojunction region across the band alignment transition regime.** Electric field: 0.136 V/nm – 0.180 V/nm

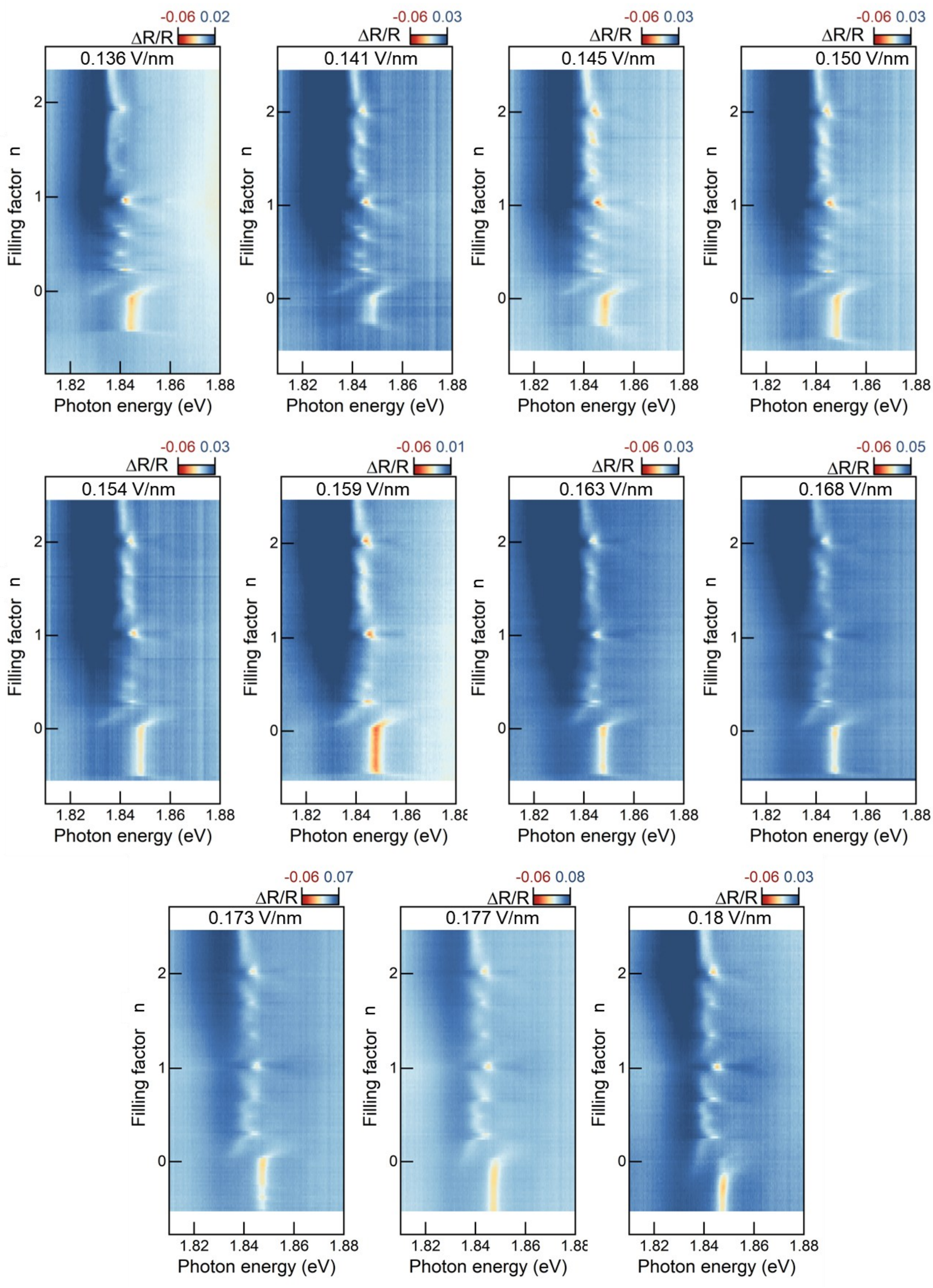

**Extended Data Fig.3 | Doping-dependent reflectance contrast spectra of $WSe_2$ 2s sensing exciton in the heterojunction region across the band alignment transition regime.** Electric field: 0.136 V/nm – 0.180 V/nm

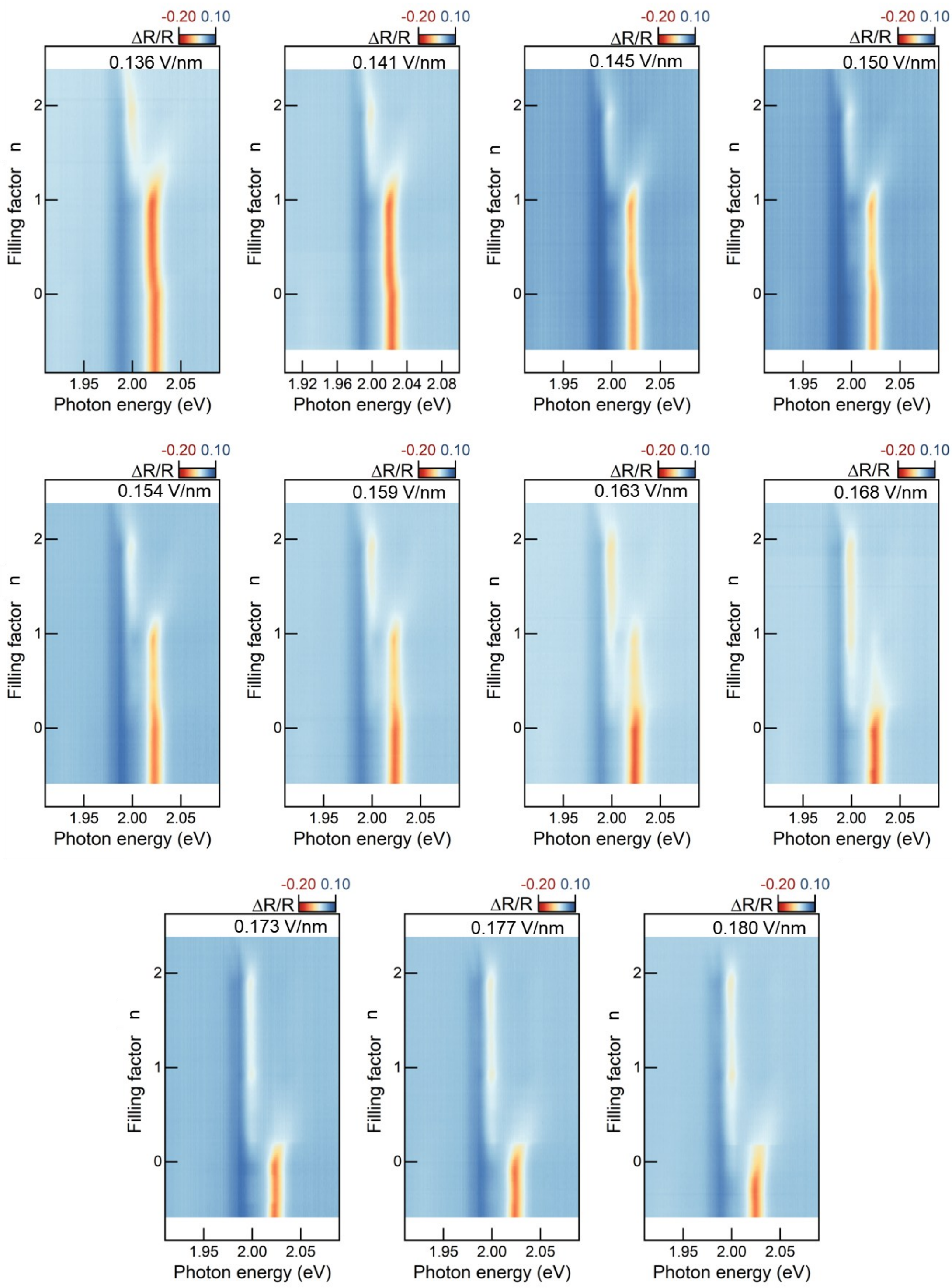

**Extended Data Fig.4 | Doping-dependent reflectance contrast spectra of $WS_2$ in the heterojunction region across band alignment transition regime.** Electric field: 0.136 V/nm – 0.180 V/nm